\documentclass{epsconf}
\usepackage{graphicx}
\usepackage{wrapfig}
\usepackage{amsmath}

\title{Modeling Vacuum Arcs}
\author{Z. Insepov$^1$, \underline{J. Norem}$^1$, T. Proslier$^1$, D. Huang$^2$, S. Mahalingam$^3$, S. Veitzer$^3$}
\institute{$^1$ Argonne National Laboratory, Argonne, IL, USA\\
$^2$ IIT, Chicago, IL USA\\
$^3$ Tech-X, Boulder, CO, USA}

\begin{document}
\maketitle

Vacuum breakdown is one of the primary limitations in the design and construction of high-energy accelerators operating with warm (copper) accelerating structures, such as muon colliders, neutrino factories, or the CLIC linear collider design  \cite{muprogram}.   Vacuum breakdown has a long history.  Starting with experiments done over 100 years ago by Earhart, Hobbs, Michelson and Millikan that first defined the process, and initial modeling by Lord Kelvin, to the present day.  An enormous number of papers have been published, exploring all the experimentally accessible variables \cite{laurant,andersbook}.  Nevertheless, the physics and the mechanisms that cause this phenomenon are still not completely understood.  There remains uncertainty about both the overall process and many of the experimental details.  To a large extent, this is due to the fact that events occur very rapidly, during which  experimental parameters vary over many orders of magnitude, and a large variety of mechanisms seem to be involved.  Among the behaviors that need  explanation is how these structures can operate for very long periods {\it without} breaking down.

\begin{wrapfigure}{r}{70mm}
\includegraphics[width=70mm]{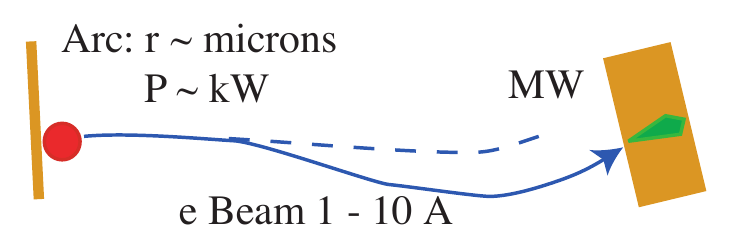}
\caption{The range of parameters of an arc in an rf cavity}
\label{fig:flow}
\end{wrapfigure}

Our interest is primarily to understand the gradient limitations of 805 and 201 MHz copper rf systems, however we find that many of the mechanisms at work in copper systems also affect superconducting rf structures.  In our copper structures arcs produce currents that short the cavities in a few rf cycles, see figure (Fig.~\ref{fig:flow}).

We assume that arcs develop as a result of mechanical failure of the surface due to electric tensile stress, ionization of fragments by field emission, and the development of a small, dense plasma that interacts with the surface primarily through self sputtering and terminates as a unipolar arc capable of producing field emitters with high enhancement factors \cite{a1,a2}.  

We have modeled the mechanisms we believe to be dominant in all stages of the arc using a number of techniques.  We use Molecular Dynamics (mechanical failure, Coulomb explosions, self sputtering), Particle-In-Cell (PIC) codes (plasma evolution), mesoscale surface thermodynamics (surface evolution), and finite element electrostatic modeling (field enhancements).  We believe this model may be more widely applicable and we are trying to constrain the physical mechanisms using data from tokamak edge plasmas, laser ablation and other environments. 

\begin{wrapfigure}{r}{60mm}
\includegraphics[width=60mm]{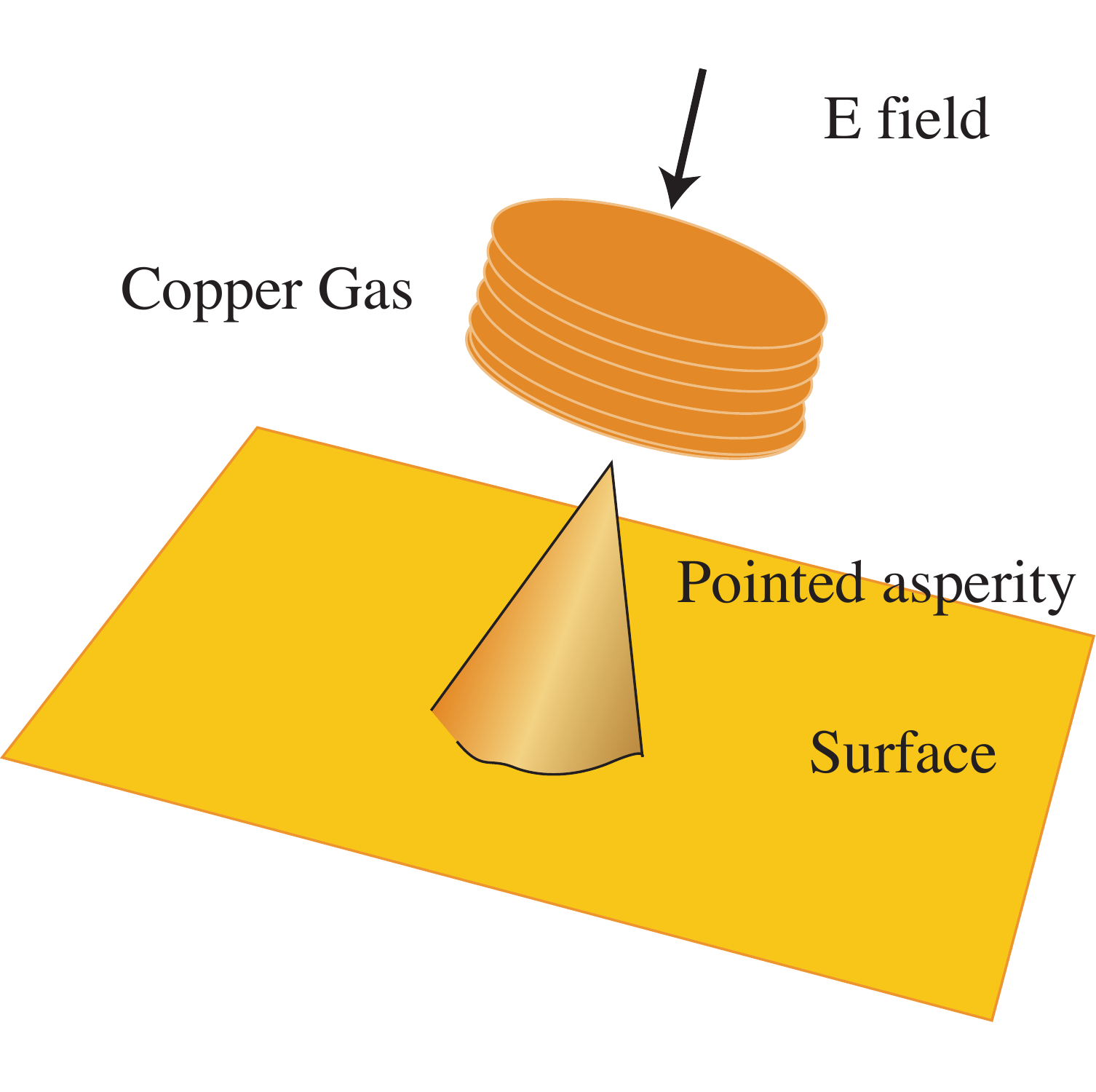}
\caption{The geometry used to model field emission ionization.}
\label{fig:flow2}
\end{wrapfigure}

\begin{wrapfigure}{r}{60mm}
\includegraphics[width=60mm]{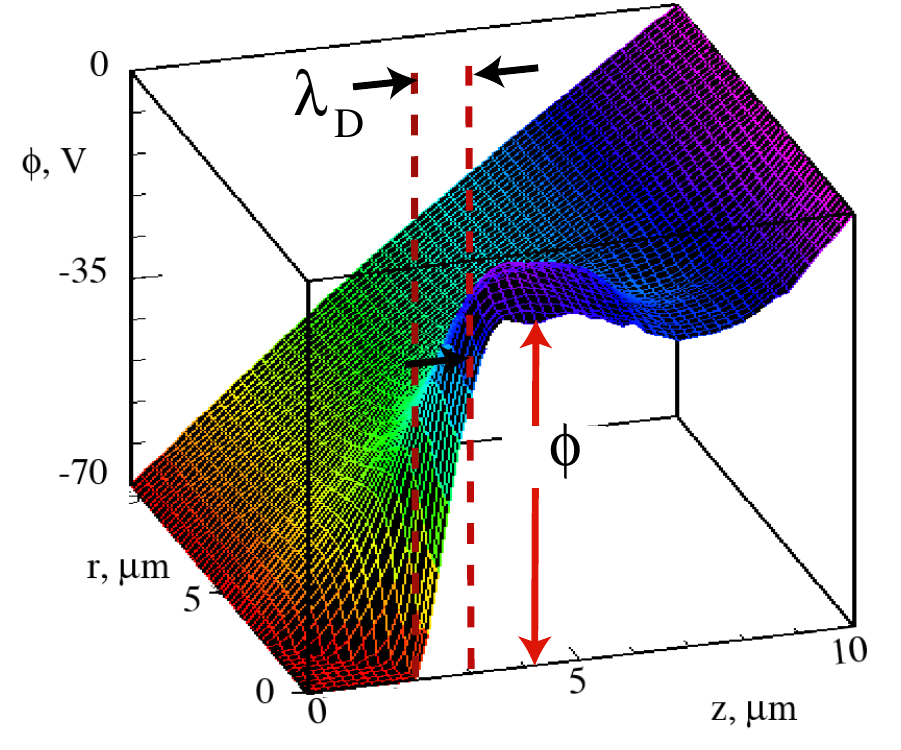}
\caption{The potential early in the discharge, with the Debye length.}
\label{fig:flow4}
\end{wrapfigure}

The initial electrostatically induced fracture is modeled using Molecular Dynamics (MD), as are the Coulomb explosion of fragments.  The initial stages of the plasma ionization, due to field emission, are modeled using OOPIC, using a geometry shown in figure (Fig.~\ref{fig:flow2}).  The potential $\phi$ in the region of the asperity during the plasma development is shown in figure (Fig.~\ref{fig:flow4}).  As the plasma develops during the initial few rf pulses, the plasma density increases roughly exponentially with time and the Debye length, $\lambda_D$, decreases, eventually to a few nm.  As the sheath potential of the plasma remains roughly constant the surface electric field continues to increase, and the granularity of the codes prevent study of maximum surface field.

Field emitted beams produce a slowly expanding, very dense, low temperature ion cloud close to the surface that enhances the local electric field on the surface, increasing the field emission and further ionization.  The stability of the plasma is primarily due to the inertial mass of tie ions, which require ns to move out of the volume of the arc under the quasi-neutrality provided by the trapped electrons.  While the electrostatic sheath potential remains roughly constant during the development of the arc, the densities of field emitted electrons, ions and trapped electrons, and the surface electric field,  increase roughly exponentially during the arc.  The code terminates, but the real limits on the plasma density are unknown, and beyond the range of the PIC code.

Our simulations show that a high density of energetic trapped electrons inside the ion cloud can be the dominant source of ionization in the early stages of the plasma growth.  These electrons oscillate in the ion potential with a period on the order of a few ps.   Figure (Fig.~\ref{fig:flow4}) shows the time development of the electron and ion populations (field emitted electrons (green) and the trapped ionization electrons (yellow) and ions (blue)).  The time constants shown in the model are compatible with experimental measurements of the rise times of arcs \cite{a1}.  

\begin{wrapfigure}{r}{100mm}
\includegraphics[width=100mm]{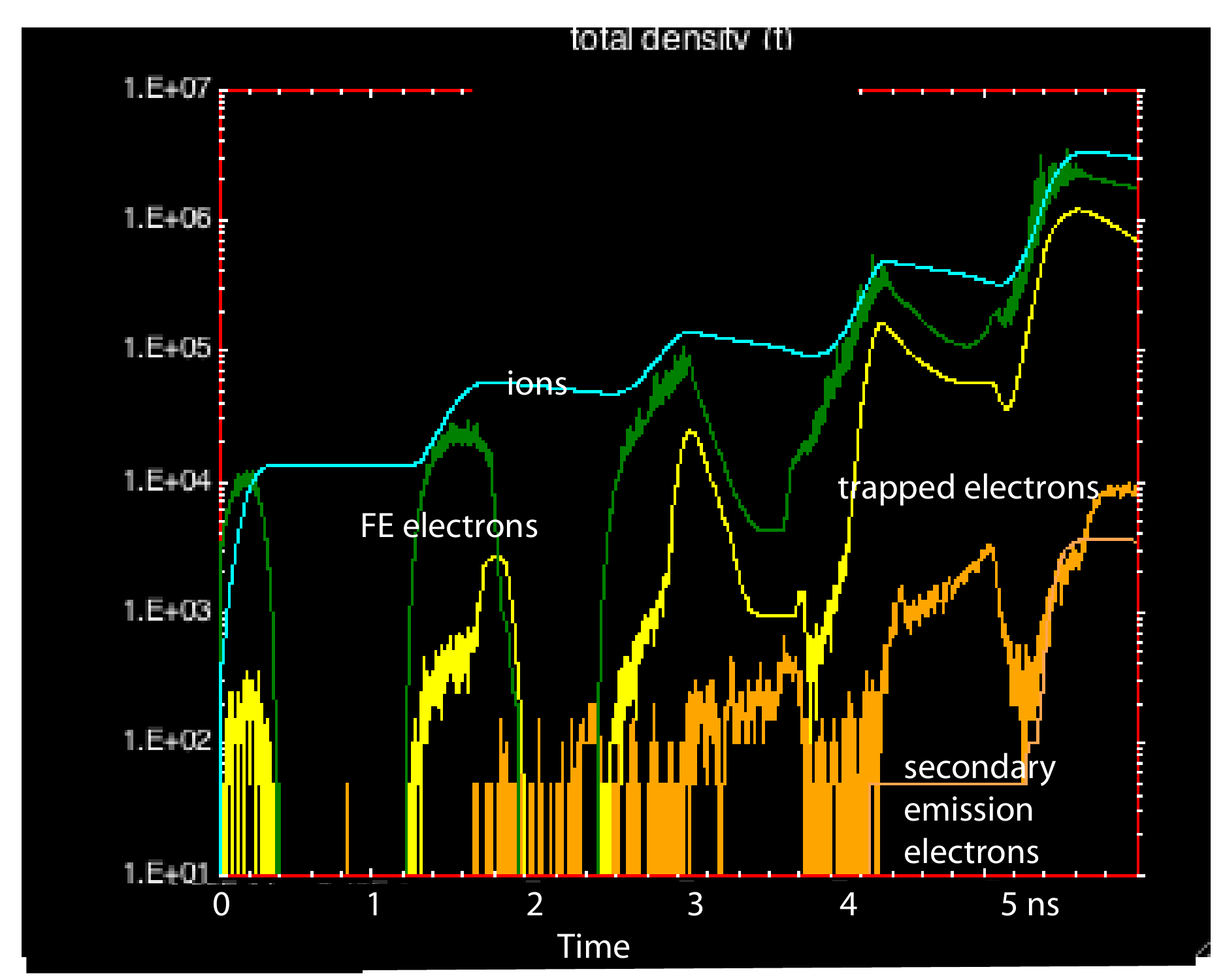}
\caption{Development of the arc plasma due to field emission}
\label{fig:flow3}
\end{wrapfigure}

Because of the high density of cold gas used in the initial state, the ion temperature at the center of the arc is roughly 1 eV, and ion energies are only significant at the arc boundaries, after acceleration by the plasma potential.  The ion density rises to 10$^{24}$ to 10$^{25}$ m$^{-3}$ as the arc develops.   Optical emission is dominated by line radiation from cold atoms, and the flux of this radiation rises with a time constant of $\sim$0.5 ns.  Continuum radiation is negligible.

The overall picture we develop is similar to the unipolar arc model of plasmas, as described by Schwirzke and others, however rf arcs exist in a oscillating potential, with the rf electric fields always sufficient to sweep electrons away from any connection to the arc in a few ps \cite{schwirzke}.   Unipolar arcs have been proposed as the primary method for surface damage and wall ablation in a wide variety of plasma environments.  These arcs seem to produce surface damage sinewhat proportional to the stored energy available to the arc, and the damage is dependent on the geometry and strength of an external magnetic field.

We have tried to associate the plasma model with surface damage produced in rf arcs.  An examination of the surface in the arc pits shows that there are small (submicron) cracks visible in places.  These cracks join to form sharp corners that we can model with COMSOL to evaluate the electrostatic enhancement factors.  We find that these enhancement factors can be larger than 100, due to the microgeometry, and we assume that the overall structure of the surface could multiply these factors by an additional term due to the local radii.  Others have seen comparably large enhancement factors on nominally clean Nb surfaces \cite{mueller2}.  While the surface areas of the individual field emitters are very small, there are a large number of them, all roughly equal, to give an effectively larger emitting area.

 \begin{wrapfigure}{r}{70mm}
\includegraphics[width=70mm]{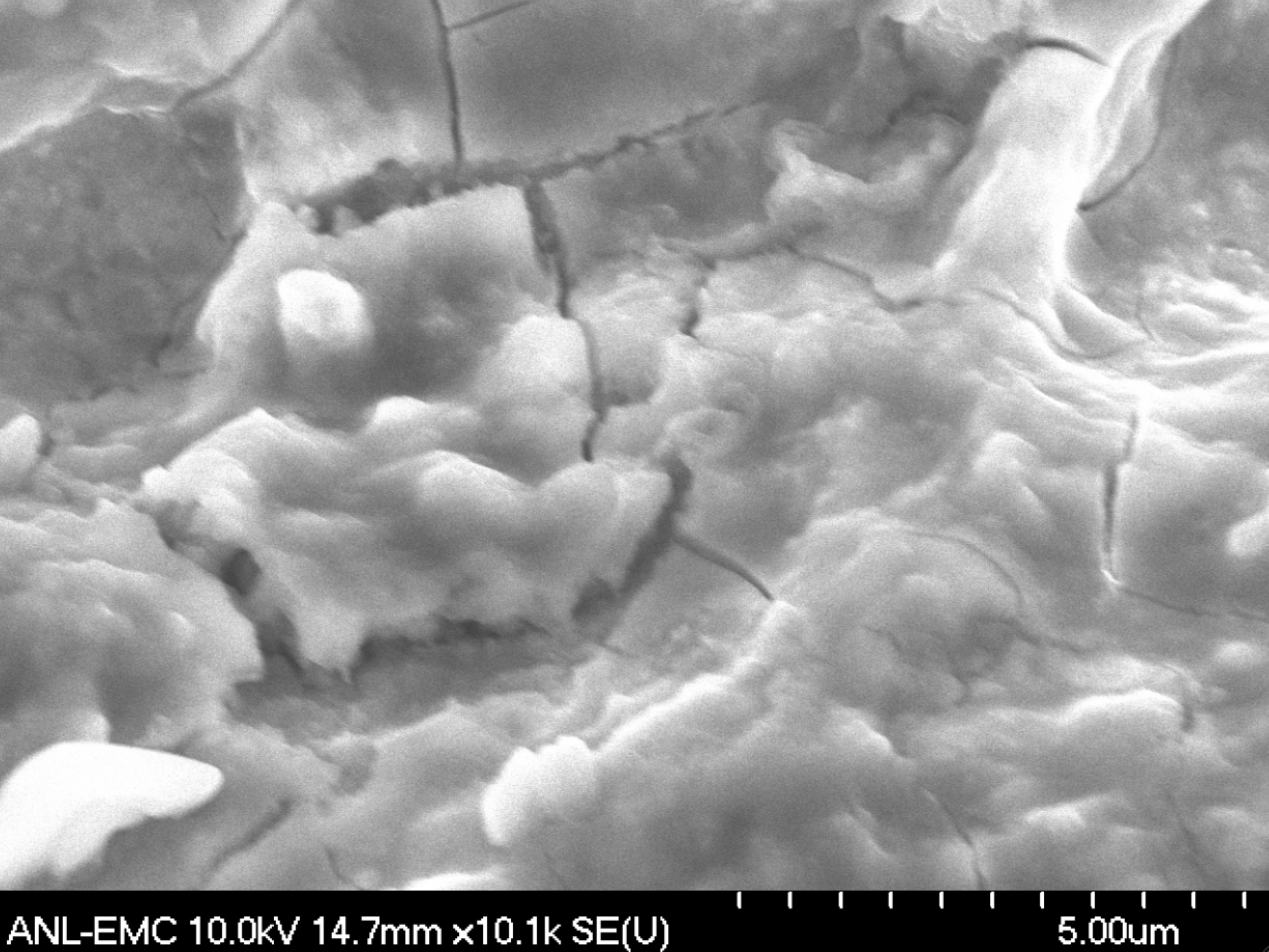}
\caption{Cracks and bubbles in arc pits.}
\label{fig:flowwww}
\end{wrapfigure}

The overall morphology of damage seen in SEM pictures of arc damage shows considerable structure (bubbles)  at dimensions of around 1 micron.  By equating the electrostatic tensile stress with the surface tension force we find that structure on this scale would exprimintally determine electric fields in the range of a few GV/m, roughly consistent with the estimates, from modeling, of sheath potentials of 70 V and Debye lengths of a few nm. At these fields the surface could field emit over large areas, producing significant current, and local magnetic fields,  ultimately shorting the plasma locally, consistent with ultimately producing a series of transient, self quenching discharges, as is seen in other data \cite{andersbook}.

This work was supported by the US/DOE Office of High Energy Physics.

\end{document}